\documentclass[aps,preprint]{revtex4}%
\usepackage{amsfonts}
\usepackage{amsmath}
\usepackage{amssymb}
\usepackage{subfigure}
\usepackage{graphicx}
\usepackage{color}%
\usepackage{graphicx}
\usepackage{float}

\usepackage{cleveref}

\begin{document}
	\title{Chaos bound of charged particles around phantom AdS black hole}
	\author{}
\author{Yue Song$^{a,b}$}
\email{syphysics@std.uestc.edu.cn}
\author{Rui Yin$^{a,c}$}
\email{yrphysics@126.com}
\author{Yiqian He$^{a,b}$}
\email{heyiqian@std.uestc.edu.cn}
\author{Benrong Mu$^{a,b,c}$}
\email{benrongmu@cdutcm.edu.cn}
\affiliation{$^{a}$ Center for Joint Quantum Studies, College of Medical Technology,
	Chengdu University of Traditional Chinese Medicine, Chengdu, 611137, PR China}
\affiliation{$^{b}$ School of Physics, University of Electronic Science and Technology of China, Chengdu, 611731, China}	
\affiliation{$^{c}$ Center for Theoretical Physics, College of Physics, Sichuan University,
	Chengdu, 610064, PR China}

	\begin{abstract}
		
In this paper, we investigated the chaos of the phantom AdS black hole in near-horizon regions and at a certain distance from the black hole, where the Lyapunov exponent was calculated by the Jacobian determinant. We found that the angular momenta of charged particles nearby the black hole affect the position of the equilibrium orbit as well as the Lyapunov exponent. When the charge is large enough and the angular momenta take particular values, the bound is violated at a certain distance from the event horizon. Concurrently, we compared the differences by analyzing tables and figures after taking different values of $\eta$, where $\eta$ is a constant indicating the nature of the electromagnetic field. When $\eta=1$, it corresponds to the Reissner-Nordstr$\ddot{o}$m-AdS black hole (RN-AdS black hole) and it represents the phantom AdS black hole when $\eta=-1$. In this way, we can draw the following conclusions. Under different values of $\eta$, the variation trends of curves related to $\lambda^{2}-\kappa^{2}$ could be different, where it can be judged whether the constraint is violated by determining if $\lambda^{2}-\kappa^{2}$ is greater than zero. In addition, the corresponding numerical values of angular momenta violating the bound in the phantom AdS black hole are much smaller than the case of RN-AdS black hole. But cases in these different black holes have the same rules to violate the bound. The large absolute values of charge and $\Lambda$ are more likely to violate the bound.
		
	\end{abstract}
	\keywords{}
	
	\maketitle
	\tableofcontents{}
	
	\bigskip{}
	
	

	\section{Introduction}
	
Further studies of black holes using holography have revealed substantial fascinating features of quantum gravity in Anti-de Sitter spacetime. These achievements make the investigations of Large-N quantum field theory be pertinent to the thermal dynamics of AdS black holes. The results indicate that some thermal Large-N unitary systems scramble the information of the local initial perturbations exponentially faster than others among the micro degrees of freedom. The above is called the fast scrambler and also black holes are presumed to be the fastest in them \cite{Banerjee:2019vff}.

Chaos is an intriguing phenomenon in strongly coupled thermal systems. On account of its sensitivity to initial conditions, chaos is a member of the unpredictable random motions. In the thermal system, the scrambling time is measured by the growth rate of $C(t)=\left\langle [V(t),W(0)]\right\rangle _{\beta}^{2}$, which can be used to better judge its chaotic behavior in the early stage. Meanwhile, the black hole as a fast scramble meets $C(t)$ $\sim$ $e^{\lambda t}$, where $\lambda $ is called as Lyapunov exponent. The Lyapunov exponent, a detector to
weigh the sensitive dependence on initial conditions, is used to measure the growth of the time-out ordered correlators (OTOCs), which is a significant probe of the chaotic behavior \cite{Shenker:2013pqa,Shenker:2014cwa,Poojary:2018esz,Gao:2022ybw}.

Recently, chaos in thermal quantum systems with multi-degree of freedom has been investigated via using an OTOC \cite{Gao:2022ybw}. On the basis of the factorization hypothesis and the assumption which supposes a large hierarchy exists between the scrambling time and the dissipation time, a conjecture that there is a general upper bound on the Lyapunov exponent is put forward by

\begin{equation}\label{eqn:Q14}
    \lambda\leq\frac{2\pi T}{\hbar}, \\
\end{equation}
where $T$ is the temperature of the system \cite{Maldacena:2015waa}. After this groundbreaking conjecture was proposed, it immediately caught people's extensive attention. A large number of chaotic phenomena have been studied in various models and theories of gravity \cite{Alishahiha:2016cjk,Asano:2015eha,Giataganas:2017guj,Gur-Ari:2015rcq,Jawad:2016kgt,Maldacena:2016hyu,Polchinski:2016xgd,Giataganas:2013dha,Huang:2017nox,Berkooz:2018jqr,Guo:2020pgq,Cardoso:2008bp,Ma:2014aha,Pradhan:2014tva}.
	
However, in recent studies, some examples according to violations of the chaos bound have been found \cite{Zhao:2018wkl,Gwak:2022xje,Kan:2021blg}. A charged test particle around the black hole can achieve static equilibrium through Lorentz force. Meanwhile, the equilibrium surface of a given black hole can be determined by the charge-to-mass ratio of the particle. In Ref. \cite{Zhao:2018wkl}, Zhao et al. considered the contributions of the sub-leading terms in the near-horizon expansion. In the meantime, the chaos bound was investigated via using effective potentials. They found that the Reissner-Nordstr$\ddot{o}$m and Reissner-Nordstr$\ddot{o}$m-AdS black holes satisfy this constraint while massive black holes such as the RN-dS black hole and black holes in Einstein-Maxwell-Dilaton, Einstein-Born-Infeld as well as Einstein-Gauss-Bonnet-Maxwell gravity violate it.

Nevertheless, the contribution of angular momentum is not considered in Ref. \cite{Zhao:2018wkl}. In practice, the Lyapunov exponent could be affected by the angular momentum of a particle. It is because that angular momentum has an influence on the effective potential which can change the magnitude of the chaotic behavior of the particle. The bound is found to be violated when considering the angular momentum and recalculating the chaotic bound in the near-horizon regions for both Reissner-Nordstr$\ddot{o}$m and Reissner-Nordstr$\ddot{o}$m-AdS black holes \cite{Lei:2021koj}.

In this paper, we investigated the effect caused by the angular momenta of charged particles on the chaos bound with circular motions of the particles around a phantom black hole. After that we studied the phantom AdS black holes. It denotes phantom AdS black holes when $\eta=-1$ and when $\eta=1$ it represents RN-AdS black holes, where $\eta$ is a constant indicating the nature of the electromagnetic field. Phantom AdS black holes have been introduced detailedly in Ref. \cite{Mo:2018hav,Quevedo:2016cge,Zhang:2017tbf}.

The Lyapunov exponent here is derived from the Jacobian determinant. Meanwhile, the position of the equilibrium orbit related to the particle is influenced by the charge-to-mass ratio and angular momentum. In order to study the effect of angular momentum, the positions of different equilibrium orbits are obtained when angular momentum is changing with the fixed charge-to-mass ratio. In this paper, numerical studies of bound are carried out in the near-horizon region as well as at a certain distance from the event horizon.

The rest of the paper is organized as follows. In Sec. \ref{sec:A}, considering the motion of a charged particle in the equatorial plane of a sphere-symmetric black hole, the general expression of Lyapunov exponent is obtained via solving the eigenvalue of Jacobian matrix. In Sec. \ref{sec:B}, we investigate that the angular momentum of a particle exerts diverse influences under different constraints of parameters on the chaotic bound in the near-horizon region as well as at a certain distance from the phantom AdS black holes. Sec. \ref{sec:C} is devoted to discussing the conclusion.

	\section{Lyapunov Exponent In A Charged Black Hole}
	\label{sec:A}
	The metric of the black hole is given by
	
\begin{equation}\label{eqn:Q1}
	ds^{2}=-F(r)dt^{2}+\frac{1}{N(r)}dr^{2}+C(r)d\theta^{2}+D(r)d\phi^{2}. \\
\end{equation}
	Meanwhile, the electromagnetic potential is shown as $A_{\mu}=A_{t}dt$. The particle with the charge $q$ has the Lagrangian as moving around the black hole, which is given by
	
\begin{equation}\label{eqn:Q2}
	\mathcal{L}=\frac{1}{2}(-F\dot{t}^{2}+\frac{\dot{r}^{2}}{N}+D\dot{\phi}^{2})-qA_{t}\dot{t}, \\
\end{equation}	
    in which $\dot{x}^{\mu}=\frac{dx^{\mu}}{d\tau}$ and $\tau$ is a suitable time. We utilize the definition of generalized momentum $\pi_{\mu}=\frac{\partial\mathcal{L}}{\partial\dot{x}}$ to obtain

\begin{equation}\label{eqn:Q3}
	\pi_{t}=-F\dot{t}-qA_{t}=-E,\qquad\pi_{r}=\frac{\dot{r}}{N},\qquad\pi_{\phi}=D\dot{\phi}=L, \\
\end{equation}	
where $E$ stands for the energy of the particle and $L$ represents the angular momentum of the particle. Therefore, the Hamiltonian for a particle is

\begin{equation}\label{eqn:Q4}
	H=\frac{-(\pi_{t}+qA_{t})^{^{2}}+\pi_{r}^{2}FN+\pi_{\phi}^{2}D^{-1}F}{2F}. \\
\end{equation}	

    By calculating the Hamiltonian, we can derive the equations of motion from it.

 \begin{equation}\label{eqn:Q5}
 	\begin{aligned}
 	&\dot{t}=\frac{\partial H}{\partial\pi_{t}}=-\frac{\pi_{t}+qA_{t}}{F},\quad\dot{\pi_{t}}=-\frac{\partial H}{\partial t}=0,\quad\ \dot{r}=\frac{\partial H}{\partial\pi_{r}}=\pi_{r}N,\\
 	&\dot{\pi_{r}}=-\frac{\partial H}{\partial r}=-\frac{1}{2}[\pi_{r}^{2}N^{'}-\frac{2qA_{t}^{'}(\pi_{t}+qA_{t})}{F}+\frac{(\pi_{t}+qA_{t})^{2}F^{'}}{F^{2}}-\pi_{\phi}^{2}D^{-2}D^{'}],\\
 	&\dot{\phi}=\frac{\partial H}{\partial\pi_{\phi}}=\frac{\pi_{\phi}}{D},\quad\dot{\pi_{\phi}}=-\frac{\partial H}{\partial\phi}=0.\\
 	\end{aligned}
 \end{equation}	

    The meaning of ''$\prime$'' in this equation is the derivative with respect to $r$. By solving the equations of motion, we obtain the derivative of radial coordinate as well as radial momentum with respect to time respectively,

\begin{equation}\label{eqn:Q6}
	\begin{aligned}
		&\frac{dr}{dt}=\frac{\dot{r}}{\dot{t}}=-\frac{\pi_{r}FN}{\pi_{t}+qA_{t}},\\
		&\frac{d\pi_{r}}{dt}=\frac{\dot{\pi_{r}}}{\dot{t}}=-qA_{t}^{'}+\frac{1}{2}[\frac{\pi_{r}^{2}FN^{'}}{\pi_{t}+qA_{t}}+\frac{(\pi_{t}+qA_{t})F^{'}}{F}-\frac{\pi_{\phi}^{2}D^{-2}D^{'}F}{\pi_{t}+qA_{t}}].\\
	\end{aligned}
\end{equation}	

To simplify the expression, we define $F_{1}=\frac{dr}{dt}$ and $F_{2}=\frac{d\pi_{r}}{dt}$. $g_{\mu\nu}\dot{x}^{\mu}\dot{x}^{\nu}=\eta$, not the same parameter compared with the $\eta$ in phantom AdS black hole, gives us a four-velocity normalization for a particle, where $\eta=0 $ for particle means a photon and $\eta=-1$ corresponds to a massive particle. The above particle is charged. We can use the normalization and Eq. $\left(\ref{eqn:Q1}\right)$ to obtain a constraint

\begin{equation}\label{eqn:Q7}
	\begin{aligned}
		&\pi_{t}+qA_{t}=-\sqrt{F(1+\pi_{r}^{2}N+\pi_{\phi}^{2}D^{-1})}.\\
	\end{aligned}
\end{equation}	

   Eq. $\left(\ref{eqn:Q6}\right)$ can be rewritten by using Eq.  $\left(\ref{eqn:Q7}\right)$ as follows

\begin{equation}\label{eqn:Q8}
	\begin{aligned}
		&F_{1}=\frac{\pi_{r}FN}{\sqrt{F(1+\pi_{r}^{2}N+\pi_{\phi}^{2}D^{-1})}},\\
		&F_{2}=-qA_{t}^{'}-\frac{1}{2}[\frac{\pi_{r}^{2}(NF)^{'}+F^{'}}{\sqrt{F(1+\pi_{r}^{2}N+\pi_{\phi}^{2}D^{-1})}}+\frac{\pi_{\phi}^{2}(D^{-1}F)^{'}}{\sqrt{F(1+\pi_{r}^{2}N+\pi_{\phi}^{2}D^{-1})}}].\\
	\end{aligned}
\end{equation}	

   The acquisition of Lyapunov is influenced by the effective potential of a particle. In the meantime, the Lyapunov exponent can be derived from the eigenvalue of a Jacobian matrix in phase space ($r$,$\pi_{r}$). The Jacobian matrix can be represented by $K_{ij}$, which can be found detailedly in Ref. \cite{Gao:2022ybw}.

   Since the particle is moving in an equilibrium orbit, $\pi_{r}=\frac{d\pi_{r}}{dt}=0$ is used to constrain the trajectory of the particle. Taking advantage of the restriction and calculating the eigenvalue, the exponent is given by

\begin{equation}\label{eqn:Q10}
	\begin{aligned}
		&\lambda^{2}=\frac{1}{4}\frac{N[F^{'}+\pi_{\phi}^{2}(D^{-1}F)^{'}]^{2}}{F(1+\pi_{\phi}^{2}D^{-1})^{2}}-\frac{1}{2}N\frac{F^{''}+\pi_{\phi}^{2}(D^{-1}F)^{''}}{1+\pi_{\phi}^{2}D^{-1}}-\frac{qA_{t}^{''}FN}{\sqrt{F(1+\pi_{\phi}^{2}D^{-1})}}.\\
	\end{aligned}
\end{equation}

   From the above equation, it is found that the exponent is influenced by angular momentum. Since angular momentum plays a significant role in Lyapunov exponent, we do not neglect angular momentum (not setting $\pi_{\phi}=0$) in this paper.

	\section{Chaos Bound And Its Violation In Phantom AdS Black Holes	}
	\label{sec:B}
	
	We delved the chaos bound by using a test particle with mass $m$ around a charged phantom AdS black hole. The force on a neutral particle is the centrifugal force, which is related to angular momentum. While the angular momentum can change the effective potential, but not enough to bring the orbits of particles closer to the horizon. When a particle has the charge $q$, changing the charge-to-mass ratio can make them close to or far from the event horizon. We investigated the effect of angular momentum on the bound in the near-horizon region as well as at a certain distance from the horizon.
	
	Before we proceed with the analysis specifically, a review of the Phantom AdS black hole should be mentioned. The action of Einstein theory with minimum coupling of the cosmological constant to the electromagnetic field \cite{Jardim:2012se} is given by

\begin{equation}\label{eqn:Q15}
		\begin{aligned}
		S=\int d^{4}x\sqrt{-g}(R+2\eta F_{\mu\nu}F^{\mu\nu}+2\Lambda).\\
		\end{aligned}
\end{equation}	

	The solution from above is obtained by
\begin{equation}\label{eqn:Q15}
	\begin{aligned}
		ds^{2}=f(r)dt^{2}-\frac{1}{f(r)}dr^{2}-r^{2}d\theta^{2}-r^{2}sin^{2}\theta d\phi^{2},\\
	\end{aligned}
\end{equation}		
where
\begin{equation}\label{eqn:Q16}
	\begin{aligned}
f(r)=1-\frac{2M}{r}-\frac{\Lambda}{3}r^{2}+\eta\frac{Q^{2}}{r^{2}}.
\end{aligned}
\end{equation}

Substituting the Eq. $\left(\ref{eqn:Q1}\right)$ into the Eq. $\left(\ref{eqn:Q15}\right)$ and the Eq. $\left(\ref{eqn:Q16}\right)$, we can obtain
	
\begin{equation}\label{eqn:Q11}
	\begin{aligned}
		&F(r)=N(r)=-1+\frac{2M}{r}+\frac{\Lambda}{3}r^{2}-\eta\frac{Q^{2}}{r^{2}},\\
		&C(r)=-r^{2},\,\, D(r)=-r^{2}sin^{2}\theta.\\
	\end{aligned}
\end{equation}	
	
	$M$ and $Q$ denote the black hole's mass and the charge of the black hole respectively. As all the parameters disappearing, this solution can be simplified to Minkowski spacetime. When $\Lambda<0$, it was considered as asymptotically anti-de Sitter. Meanwhile, $\eta=1$ can make an Reissner-Nordstr$\ddot{o}$m-AdS (RN-AdS) black hole recover. If $f(r)=0$ is solved, two effective positive roots can be obtained when $\eta=1$ while only one effective positive root can be obtained when $\eta=-1$. The electromagnetic potential is $\frac{Q}{r_{+}}$. The event horizon ($r_{+}$) is the positive solution of $f(r)=0$. In the equatorial plane, we obtain $\theta=\frac{\pi}{2}$ and then $D(r)=-r^{2}$.
	
To begin with, the positions of equilibrium orbits need finding. Then the formula for surface gravity is as follow
	
\begin{equation}\label{eqn:Q13}
		\begin{aligned}
			&\kappa=\frac{1}{2}\frac{\partial F}{\partial r}\mid_{r=r_{+}}=-\frac{M}{r_{+}^{2}}+\frac{Q^{2}\eta}{r_{+}^{3}}+\frac{r_{+}\Lambda}{3}.\\
		\end{aligned}
\end{equation}

Combining the constraint $\pi_{r}=\frac{d\pi_{r}}{dt}=0$ with Eq. $\left(\ref{eqn:Q11}\right)$ and Eq. $\left(\ref{eqn:Q13}\right)$, we obtained an expression for the radii of equilibrium orbits. Because of the complexity, we decided to use the numerical analysis about the position $r_{0}$ of equilibrium orbit, which simplifies the calculations. We set $M=1$, $m=1$ and $q=15$. And then the positions of certain specific orbits in \Cref{tab:01,tab:02,tab:03,tab:04,tab:05} were obtained.
		
\begin{figure}[H]
	\begin{center}
		\subfigure[{The figures of $\lambda^{2}-\kappa^{2}$ with respect to $L$ and $Q$ when $\Lambda=-0.5$, $\eta=1$ and $q=15$.}]{
			\includegraphics[width=0.8\textwidth]{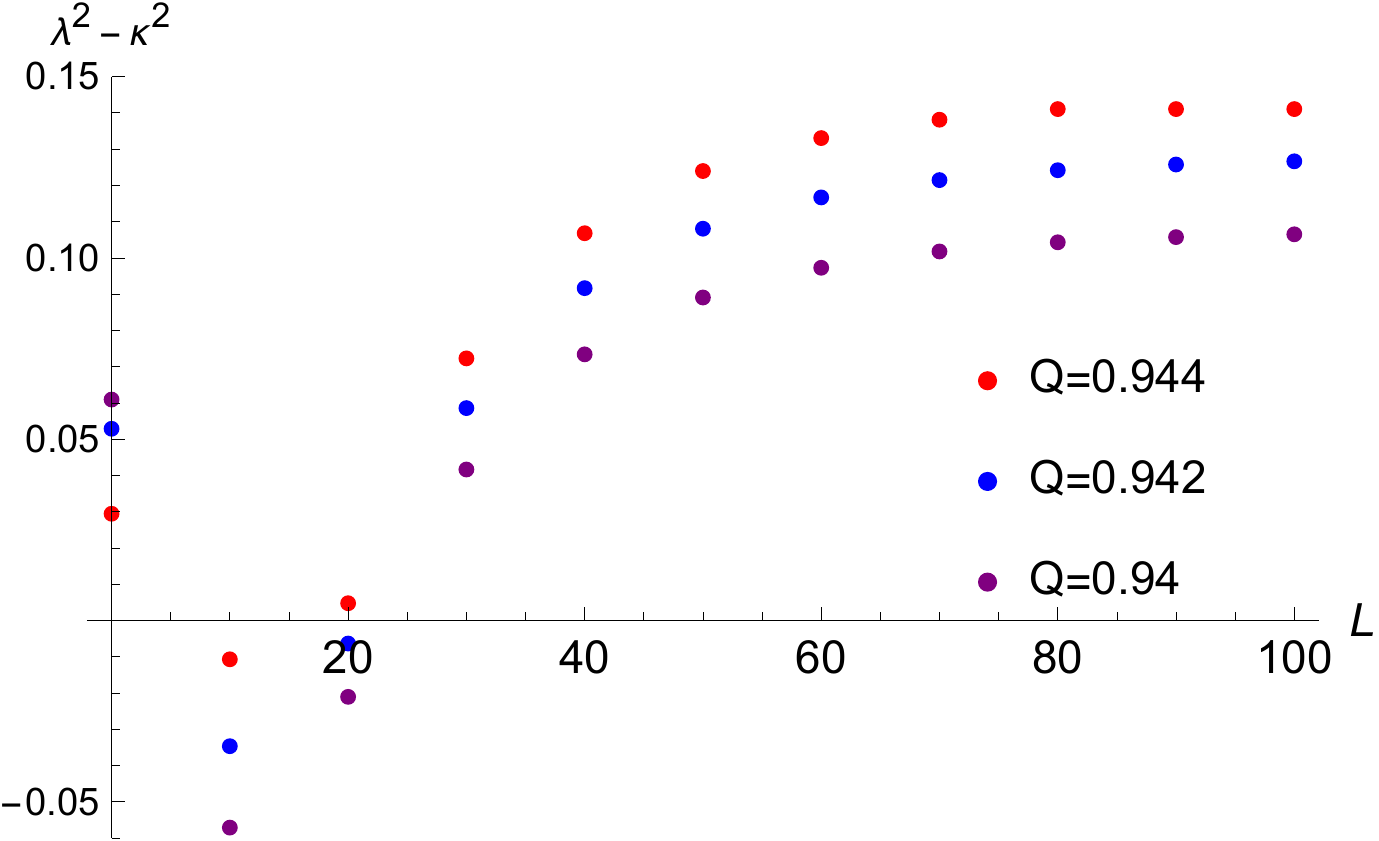}\label{fig:n1Q}}
		\subfigure[{The figures of $\lambda^{2}-\kappa^{2}$ with respect to $L$ and $Q$ when $\Lambda=-0.5$, $\eta=-1$ and $q=15$.}]{
			\includegraphics[width=0.8\textwidth]{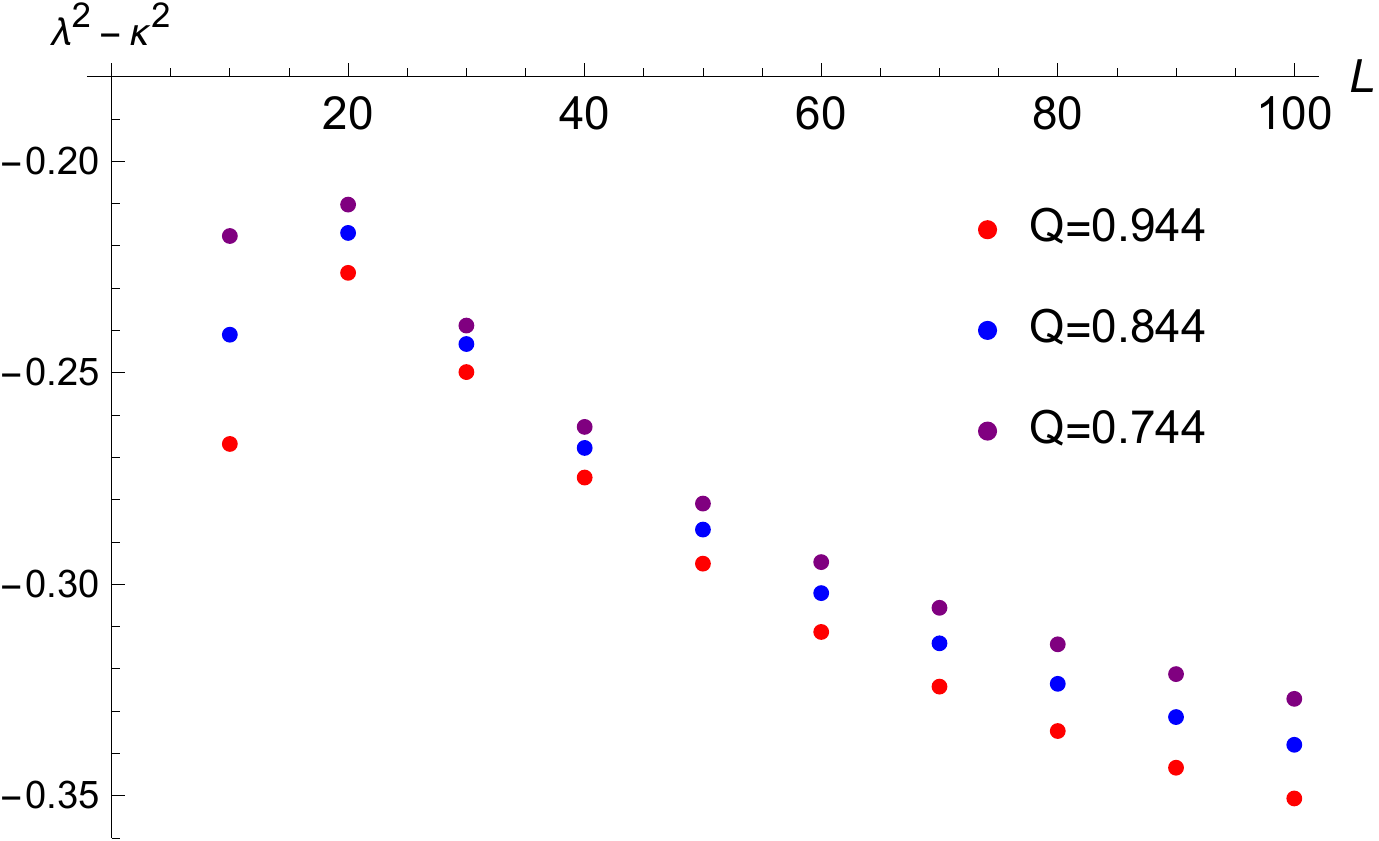}\label{fig:nf1Q}}
	\end{center}
	\caption{$\lambda^{2}-\kappa^{2}$ for different amounts of charge.}%
	\label{fig:01}
\end{figure}

\begin{figure}[H]
	\begin{center}
		\subfigure[{The figures of $\lambda^{2}-\kappa^{2}$ with respect to $L$ and $\Lambda$ when $Q=0.944$, $\eta=1$ and $q=15$.}]{
			\includegraphics[width=0.8\textwidth]{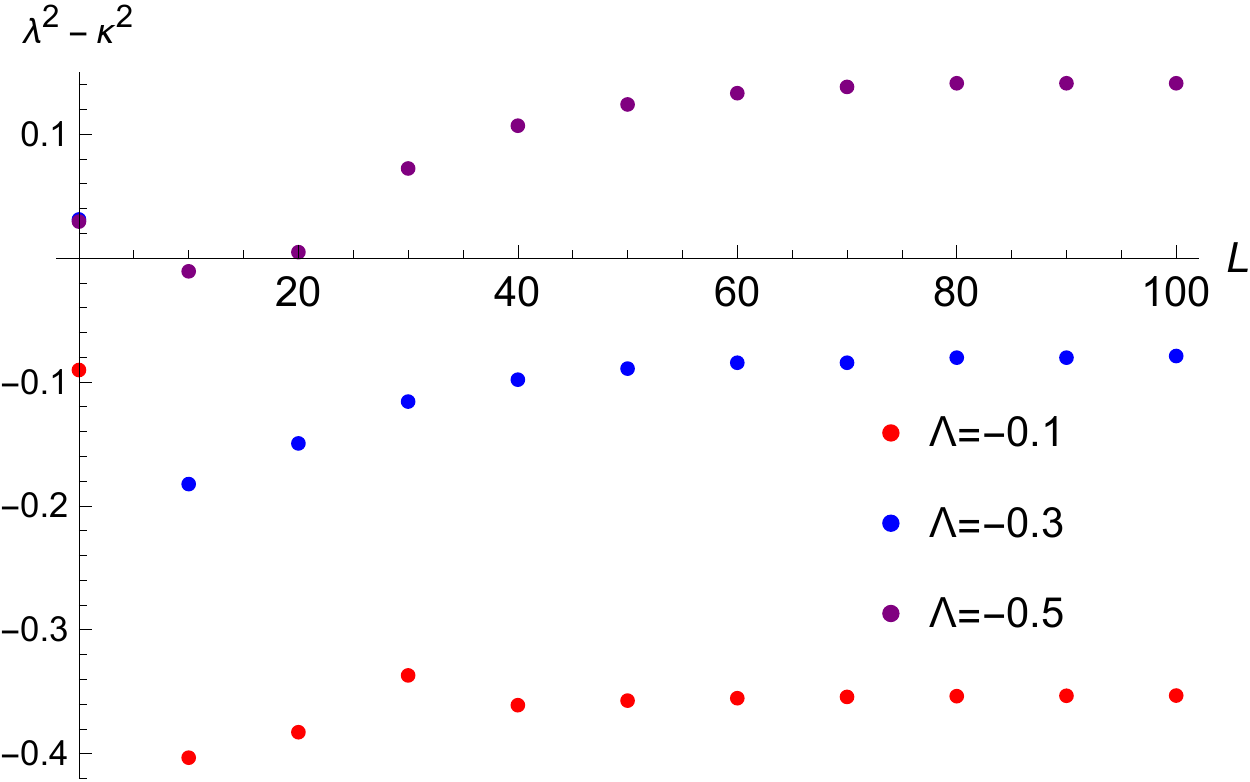}\label{fig:n1A}}
		\subfigure[{The figures of $\lambda^{2}-\kappa^{2}$ with respect to $L$ and $\Lambda$ when $Q=0.944$, $\eta=-1$ and $q=15$.}]{
			\includegraphics[width=0.8\textwidth]{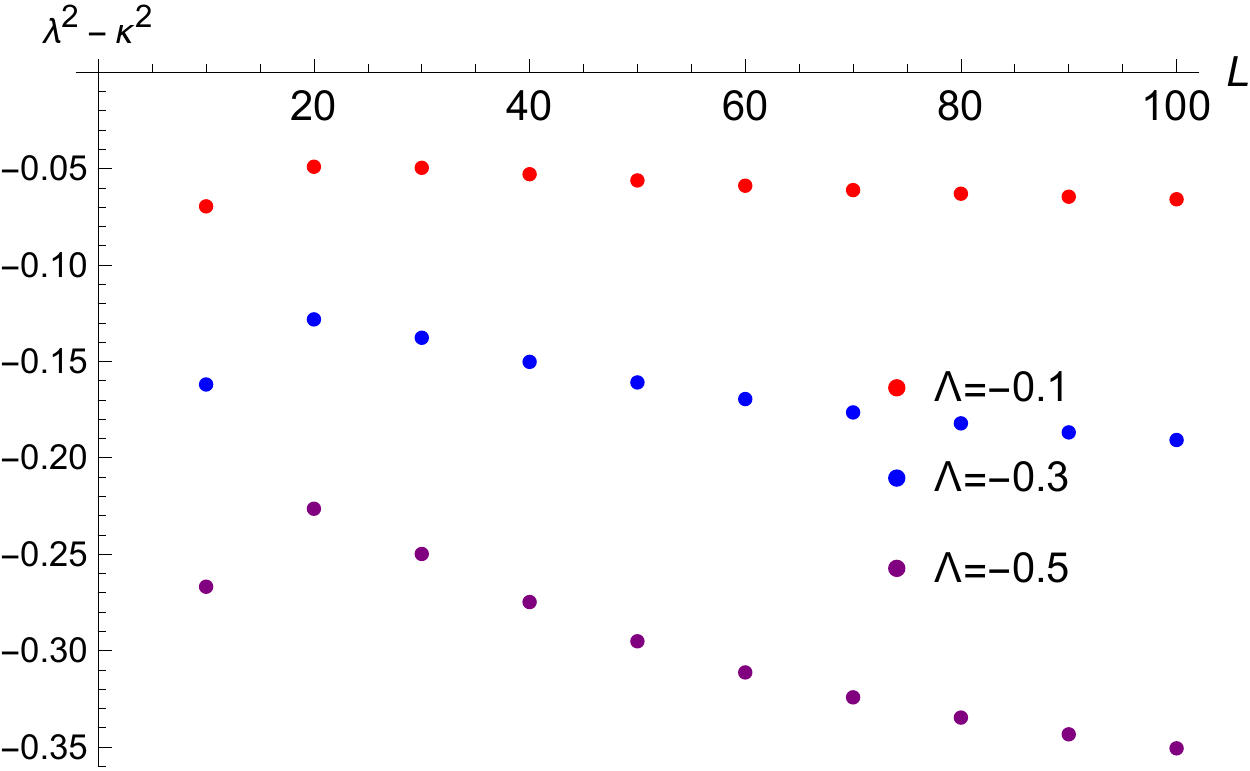}\label{fig:nf1A}}
	\end{center}
	\caption{$\lambda^{2}-\kappa^{2}$ for different amounts of $\Lambda$. }%
	\label{fig:02}
\end{figure}

\begin{figure}[H]
	\begin{center}
		\subfigure[{The figures of $\frac{\lambda^{2}-\kappa^{2}}{|\lambda^{2}-\kappa^{2}|}lg[|\lambda^{2}-\kappa^{2}|]$ with respect to $L$ and $Q$ when $\Lambda=-0.5$, $\eta=-1$ and $q=15$.}]{
			\includegraphics[width=0.8\textwidth]{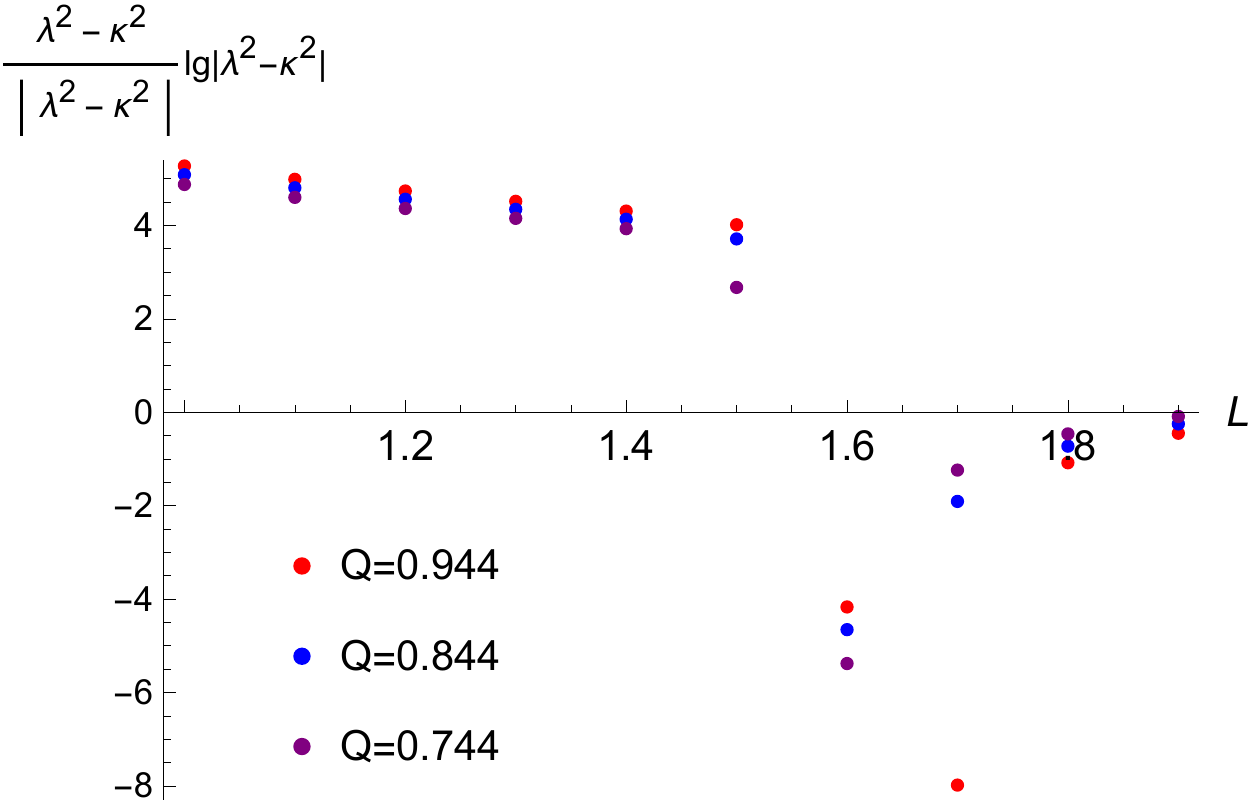}\label{fig:nf1Qx}}
		\subfigure[{The figures of $\frac{\lambda^{2}-\kappa^{2}}{|\lambda^{2}-\kappa^{2}|}lg[|\lambda^{2}-\kappa^{2}|]$ with respect to $L$ and $\Lambda$ when $Q=0.944$, $\eta=-1$ and $q=15$.}]{
			\includegraphics[width=0.8\textwidth]{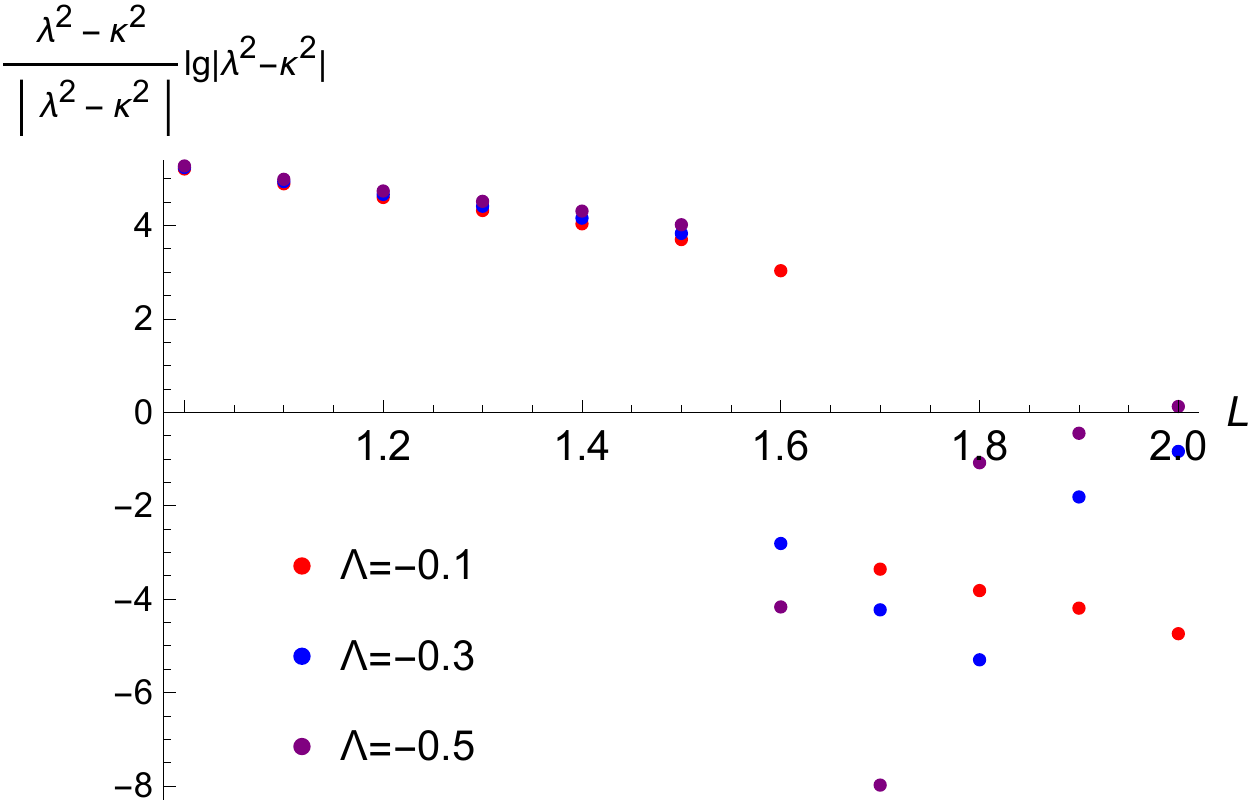}\label{fig:nf1Ax}}
	\end{center}
	\caption{$\frac{\lambda^{2}-\kappa^{2}}{|\lambda^{2}-\kappa^{2}|}lg[|\lambda^{2}-\kappa^{2}|]$ for different amounts of $\Lambda$ and $Q$. }%
	\label{fig:03}
\end{figure}

As for the figures, $\lambda^{2}-\kappa^{2}$ was obtained by using Eq. $\left(\ref{eqn:Q10}\right)$ and Eq. $\left(\ref{eqn:Q13}\right)$. Fig. \ref{fig:01} presents the influence of charge in cases of different $\eta$. Fig. \ref{fig:02} shows the influence of $\Lambda$ with different $\eta$. Then we discovered that the specific angular momenta with $\eta=-1$ violating the bound do not satisfy the definition fields of Fig. \ref{fig:01} and Fig. \ref{fig:02}. Thus, we calculated and plotted Fig. \ref{fig:03} to exhibit the situation of violation in the case of $\eta=-1$ purposely. The values of $\lambda^{2}-\kappa^{2}$ vary with angular momenta drastically, which means it is hard to be presented in one figure. Therefore, we adopted $\frac{\lambda^{2}-\kappa^{2}}{|\lambda^{2}-\kappa^{2}|}lg[|\lambda^{2}-\kappa^{2}|]$ to make the Fig. \ref{fig:03} more clearly. Particularly, the values of $|\lambda^{2}-\kappa^{2}|$ in Fig. \ref{fig:03} are larger than one, which indicates the sign of $\frac{\lambda^{2}-\kappa^{2}}{|\lambda^{2}-\kappa^{2}|}lg[|\lambda^{2}-\kappa^{2}|]$ is only up to the sign of $\lambda^{2}-\kappa^{2}$.

\begin{table}[htb]
	\begin{tabular}{|l|l|l|l|l|l|l|l|}
		\hline
		&$L$ & 1 & 2 & 3 & 5 & 10 & 20 \\ \hline
		$r_{0}$	& $\eta=1$ &0.84355  &0.84388  &0.84446  & 0.84655 & 0.86636 & 1.11622 \\ \cline{2-8}
		& $\eta=-1$ & 1.00416 & 1.70796 & 1.73342 & 1.80041 & 1.98734 & 2.28241 \\ \hline
	\end{tabular}
	\caption{{When $M=1$, $Q=0.944$, $\varLambda=-0.5$ and $q=15$, the position of the equilibrium orbit varies with the value of the angular momentum. Meanwhile, the event horizon is located at $r_{+}=0.79173$ when $\eta=1$. When $\eta=-1$, the event horizon is located at $r_{+}=1.70195$.}}
	\label{tab:01}
\end{table}

\begin{table}[htb]
	\begin{tabular}{|l|l|l|l|l|l|l|l|}
		\hline
		&$L$ & 1 & 2 & 3 & 5 & 10 & 20 \\ \hline
		$r_{0}$	& $\eta=1$ &0.77010  &0.77071  &0.77179  &0.77582 & 0.81898 & 1.09788 \\ \cline{2-8}
		& $\eta=-1$ & 1.00387  &1.49177 & 1.52126 & 1.59621 &  1.79472 &  2.09879 \\ \hline
	\end{tabular}
	\caption{{When $M=1$, $Q=0.911$, $\varLambda=-1$ and $q$ = 15, the position of the equilibrium orbit varies with the value of the angular momentum. Meanwhile, the event horizon is located at $r_{+}=0.69968$ when $\eta=1$. When $\eta=-1$, the event horizon is located at $r_{+}=1.48004$.}}
	\label{tab:02}
\end{table}

\begin{table}[htb]
	\begin{tabular}{|l|l|l|l|l|l|l|l|}
		\hline
		&$L$ & 1 & 2 & 3 & 5 & 10 & 20 \\ \hline
		$r_{0}$	& $\eta=1$ &0.69500  &0.69527  &0.69575  &0.69771& 0.75494 & 1.07362\\ \cline{2-8}
		& $\eta=-1$ &1.00349&1.36685 &1.39887 & 1.47826 &1.68171 &1.98792 \\ \hline
	\end{tabular}
	\caption{{When $M$ = 1, $Q$ = 0.889, $\varLambda$ = -1.5 and $q$ = 15, the position of the equilibrium orbit varies with the value of the angular momentum. Meanwhile, the event horizon is located at $r_{+}$=0.66961 when $\eta=1$. When $\eta=-1$, the event horizon is located at $r_{+}$=1.35126.}}
	\label{tab:03}
\end{table}	

Analyzing the tables and figures, we have come to the following conclusions.

 From \Cref{tab:01,tab:02,tab:03}, it can be clearly seen that when $\eta=1$, $r_{0}$ and angular momentum are positively correlated and also $r_{0}$ is always larger than the radius of the horizon. When $\eta=-1$, $r_{0}$ and angular momentum also have a positive correlation, but $r_{0}$ is smaller than the radius of horizon at first and later larger than the radius of horizon. From Table \ref{tab:04}, it can be found that when other factors are consistent, the larger the absolute value of $\Lambda$ is, the smaller $r_{0}$ is, no matter which black hole it is. As for Table {\ref{tab:05}}, when $\eta=1$, we found that $r_{0}$ gradually decreases with the increase of $Q$ when other influencing factors remain unchanged.

 When the cosmological constant and the values of angular momenta are fixed, different values of charges produce different magnitudes of $\lambda^{2}-\kappa^{2}$. According to Fig. \ref{fig:01}, as for $\eta=1$, the high magnitude of charge make it more likely to violate the bound. While $\eta=-1$, the smaller the value of charge is, the closer $\lambda^{2}-\kappa^{2}$ is to zero. According to Fig. \ref{fig:02}, as for $\eta=1$, the large absolute value of $\Lambda$ make it more likely to violate the bound. While for $\eta=-1$, the smaller absolute value of $\Lambda$ is, the closer $\lambda^{2}-\kappa^{2}$ is to zero.

 The angular momentum of the particle affects the bound when the numerical values of other parameters are fixed. No matter what magnitude $\eta$ is chosen, certain particular values of angular momenta violate the bound with the enough charge. According to the Fig. \ref{fig:02} and Fig. \ref{fig:03}, the values of angular momenta with $\eta=1$ violating the bound are much larger than the values of corresponding angular momenta with $\eta=-1$.

 In Fig. \ref{fig:03}, when $\eta=-1$, the large value of charge has higher feasibility to cause the violation with the fixed cosmological constant and angular momentum. If the values of charge and angular momentum are fixed, the large absolute value of $\Lambda$ has higher feasibility to cause the violation.

\begin{table}[htb]
	\begin{tabular}{|l|ll|l|l|l|l|l|l|}
		\hline
		&           $L$         &  & 1 & 2 & 3 & 5 & 10 & 20   \\ \hline
		& \multicolumn{1}{l|}{} & $\varLambda$=-0.5 & 1.00012 & 1.05880 & 1.06498 & 1.08444 & 1.16617 & 1.36632 \\ \cline{3-9}
		& \multicolumn{1}{l|}{$\eta=1$} & $\varLambda$=-1 & 0.87311 & 0.87579 & 0.88034 & 0.89553 & 0.97304 & 1.19902  \\ \cline{3-9}
		$r_{0}$	& \multicolumn{1}{l|}{} & $\varLambda$=-1.5 & 0.69500 & 0.69527 & 0.69575 & 0.69771 & 0.75494 & 1.07362 \\ \cline{2-9}
		& \multicolumn{1}{l|}{} & $\varLambda$=-0.5 & 1.00441 & 1.68699 & 1.71431 & 1.78527 & 1.97856 & 2.27611 \\ \cline{3-9}
		& \multicolumn{1}{l|}{$\eta=-1$} & $\varLambda$=-1 & 1.00395 & 1.48473 & 1.51504 & 1.59163 & 1.79265 & 2.09783 \\ \cline{3-9}
		& \multicolumn{1}{l|}{} & $\varLambda$=-1.5 & 1.00349 & 1.36685 & 1.39887 & 1.47826 & 1.68171 & 1.98792 \\ \hline
\end{tabular}
	\caption{{When $M=1$, $Q=0.889$ and $q=15$, the position of the equilibrium orbit varies with the value of the angular momentum. When $\eta=1$ and $\varLambda=-0.5$, $-1$ and $-1.5$, the radius of the event horizon $r_{+}$ is $0.56041$, $0.58784$ and $0.66961$ respectively. When $\eta=-1$ and $\varLambda=-0.5$, $-1$ and $-1.5$, the radius of the event horizon $r_{+}$ is $1.68006$, $1.47249$ and $1.35126$ respectively.}}
	\label{tab:04}
\end{table}

\begin{table}[htb]
	\begin{tabular}{|l|ll|l|l|l|l|l|l|}
		\hline
		&           $L$         &  & 1 & 2 & 3 & 5 & 10 & 20   \\ \hline
		& \multicolumn{1}{l|}{} & $Q$=0.94 & 0.88898 & 0.88991 & 0.89153 & 0.89716 & 0.93513 & 1.14835 \\ \cline{3-9}
		& \multicolumn{1}{l|}{$\eta=1$} & $Q$=0.942 & 0.87145 & 0.87215 & 0.87336 & 0.87764 & 0.90959 & 1.13319  \\ \cline{3-9}
		$r_{0}$	& \multicolumn{1}{l|}{} & $Q$=0.944 & 0.84355 & 0.84388 & 0.84446 & 0.84655 & 0.86636 & 1.11622 \\ \cline{2-9}
		& \multicolumn{1}{l|}{} & $Q$=0.94 & 1.00418 & 1.70641 & 1.73200 & 1.79926 & 1.98664 & 2.28188 \\ \cline{3-9}
		& \multicolumn{1}{l|}{$\eta=-1$} & $Q$=0.942 & 1.00417 & 1.70719 & 1.73271 & 1.79984 & 1.98699 & 2.28215 \\ \cline{3-9}
		& \multicolumn{1}{l|}{} & $Q$=0.944 & 1.00416 & 1.70796 & 1.73342 & 1.80041 & 1.98734 & 2.28241 \\ \hline
	\end{tabular}
	\caption{{When $M=1$, $\varLambda=-0.5$ and $q=15$, the position of the equilibrium orbit varies with the value of the angular momentum. When $\eta=1$ and $Q=0.94$, $0.942$ and $0.944$, the radius of the event horizon $r_{+}$ is $0.74486$, $0.76311$ and $0.79173$ respectively. When $\eta=-1$ and $Q=0.94$, $0.942$ and $0.944$, the radius of the event horizon $r_{+}$ is $1.70034$, $1.70114$ and $1.70195$ respectively.}}
	\label{tab:05}
\end{table}

	\section{Conclusions}
	\label{sec:C}
	In this paper, we studied the chaotic bound of a phantom black hole in the near-horizon region and at a certain distance from the event horizon. Situations in different values of $\eta$ were compared. The angular momentum of charged particle plays an important role in the investigation, which affects not only the value of Lyapunov exponent, but also the position of equilibrium orbit. By fixing the charge-to-mass ratio and changing the value of the angular momentum, we obtained the positions of equilibrium orbits. Meanwhile, $r_{0}$ always has a positive correlation with angular momentum regardless of the value of $\eta$. And we also learned that the value of $r_{0}$ decreases as the absolute value of $\Lambda$ increases.
	
	Then we analyzed the figures which show that the bound is violated at a certain distance from the horizon when the charge is large enough and the angular momenta take particular values. Further, in the two cases of different black holes, it can be seen from the figures that the conclusions are much different. For instance, the values of specific angular momenta with $\eta=1$ violating the bound are much larger than the values of corresponding angular momenta in $\eta=-1$. Via comparing Fig.\ref{fig:01} with Fig.\ref{fig:03}, it is found that the variation trends of $\lambda^{2}-\kappa^{2}$ in two cases of black holes are entirely different when other parameters are fixed. Nevertheless, $\lambda$ with different $\eta$ has the same rule to violate the bound, which is as follow. The large absolute values of charge and $\Lambda$ are more likely to violate the bound.

\begin{acknowledgments}
	We are grateful to  Deyou Chen,Jing Liang, Peng Wang, Haitang Yang, Jun Tao and Xiaobo Guo for useful discussions. The authors contributed equally to this work. This work is supported in part by NSFC (Grant No. 11747171), Xinglin Scholars Project of Chengdu University of Traditional Chinese Medicine (Grant no.QNXZ2018050).
\end{acknowledgments}

\end{document}